\documentclass[dvips]{article}
\usepackage{icrctc07}

\newcommand{\la}{\,\rlap{\raise 0.5ex\hbox{$<$}}{\lower 1.0ex\hbox{$\sim$}}\,}
\newcommand{\ga}{\,\rlap{\raise 0.5ex\hbox{$>$}}{\lower 1.0ex\hbox{$\sim$}}\,}

\title{Ultrahigh energy cosmic rays as heavy nuclei from cluster accretion shocks}
\shorttitle{UHECRs as heavy nuclei from clusters}
\authors{Susumu Inoue$^{1}$, G\"unter Sigl$^{2,3}$, Francesco Miniati$^{4}$, Eric Armengaud$^{5}$}
\shortauthors{S. Inoue et al}
\afiliations
{$^1$National Astronomical Observatory of Japan, Osawa 2-21-1, Mitaka, 181-8588 Tokyo, Japan\\
$^2$AstroParticules et Cosmologie, 10 rue Alice Domon et L\'eonie Duquet, 75205 Paris, France \\
$^3$Institut d'Astrophysique de Paris, 98 bis boulevard Arago, 75014 Paris, France\\
$^4$Physics Department, ETH Z\"urich, 8093 Z\"urich, Switzerland\\
$^5$DSM/DAPNIA, CEA/Saclay, 91191, Gif-sur-Yvette, France}
\email{inoue@th.nao.ac.jp}

\abstract
{Large-scale accretion shocks around massive clusters of galaxies,
generically expected in hierarchical scenarios of cosmological structure formation,
are shown to be potential sources of the observed ultrahigh energy cosmic rays (UHECRs)
by accelerating a mixture of heavy nuclei including the iron group elements.
Current observations can be explained if the source composition at injection for the heavier nuclei
is somewhat enhanced from simple expectations for the accreting gas.
The proposed picture should be testable by current and upcoming facilities in the near future
through characteristic features in the UHECR spectrum, composition and anisotropy.
The associated X-ray and gamma-ray signatures are also briefly discussed.}

\begin{document}
\maketitle

\section{Introduction}

The origin of UHECRs with energies $10^{18}$-$10^{20}$ eV and above
remains one of the biggest mysteries in physics and astrophysics \cite{nw00}.
Only a few types of astrophysical objects appear capable of accelerating UHECRs to the highest observed energies,
such as the jets of radio-loud active galactic nuclei or gamma-ray bursts \cite{hil84, ino07}.
However, no unambiguous identification with any kind of source has been achieved so far.

In the currently favored picture of hierarchical structure formation in the CDM cosmology,
all massive clusters of galaxies should be surrounded by strong accretion shocks,
as a consequence of continuing infall of dark matter and baryonic gas \cite{ryu03}.
Such shocks should be interesting sites of particle acceleration,
and have also been proposed as sources of UHECRs \cite{nma95}.
Here we summarize our recent work on this subject invoking UHECR nuclei;
more details can be found in Ref. \cite{isma07}.

\section{Model}

For clusters of mass $M$, the rate of gas kinetic energy dissipation through accretion shocks
can be estimated as
$L_{\rm acc} \simeq 9 \times 10^{45} (M/{10^{15} M_\odot})^{5/3} {\rm erg\ s^{-1}}$ \cite{kwl04}.
Taking the number density of clusters with $M \ga 10^{15} M_\odot$ to be
$n_s=2 \times 10^{-6} {\rm Mpc^{-3}}$ as observed locally within 120 Mpc \cite{rb02},
we can evaluate the total power density from cluster accretion to be
$P_{\rm acc} \sim L_{\rm acc} n_s = 2 \times 10^{40} {\rm erg\ s^{-1} Mpc^{-3}}$,
which should reasonably accommodate the energy budget of UHECRs.

\begin{figure}[htb!]
\begin{center}
\includegraphics [width=0.47\textwidth]{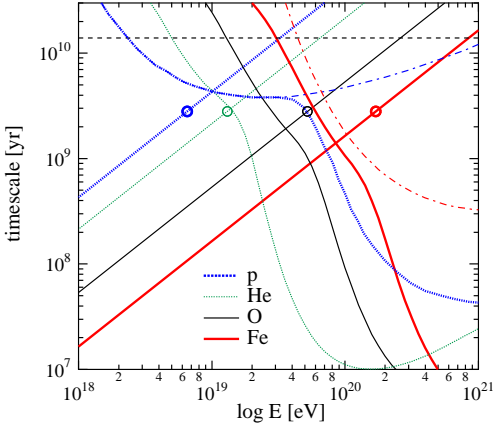}
\end{center}
\caption{
Comparison of timescales at cluster accretion shocks
for shock acceleration $t_{\rm acc}$ (diagnonal lines),
and energy losses from interactions with background radiation fields (curves),
for protons (thick dotted), He (thin dotted), O (thin solid) and Fe nuclei (thick solid).
The photopair timescales are denoted separately for p and Fe (dot-dashed).
Also indicated are the Hubble time $t_H$ (dashed)
and the escape-limited $E_{\max}$ (circles).
}\label{fig:tacc}
\end{figure}

However, estimates of the maximum energy $E_{\max}$ for protons
seem to fall short of $10^{20}$ eV by 1-2 orders of magnitude \cite{krj96, ias05}.
A fiducial cluster of $M=2 \times 10^{15} M_\odot$
has shock radius $R_s \simeq 3.2$ Mpc
and shock velocity $V_s=(4/3)(GM/R_s)^{1/2} \simeq 2200$ km/s.
The shock magnetic field is taken to be $B_s=1 \mu$G, as suggested by some recent observations \cite{fn06}.
The timescale for shock acceleration of particles with energy $E$ and charge $Z$
is $t_{\rm acc}=20 \kappa(E)/V_s^2 = (20/3) (E c/Z e B_s V_s^2)$,
assuming the Bohm limit for the diffusion coefficient  $\kappa(E)$,
which can be induced by CR wave excitation \cite{bel04}
and is compatible with observations of supernova remnant shocks \cite{par06}.
To be compared are the energy loss timescales
for photopair and photopion interactions with the cosmic microwave background (CMB) \cite{ss99},
the escape time from the acceleration region $t_{\rm esc} \sim R_s^2/5 \kappa(E)$ \cite{rb93},
and the Hubble time $t_H$.
As is clear in Fig.\ref{fig:tacc}, for protons $E_{\max} \sim 10^{18}$-$10^{19}$ eV, confirming previous findings.

On the other hand, heavy nuclei with higher $Z$ have correspondingly shorter $t_{acc}$,
and Fe may be accelerated up to $10^{20}$ eV in the same conditions,
notwithstanding energy losses by photodisintegration and photopair interactions
with the far infrared background (FIRB) and CMB (Fig.\ref{fig:tacc}).
In order to explore whether nuclei from cluster accretion shocks
can provide a viable picture of UHECR origin,
detailed propagation calculations of UHE nuclei above $10^{19}$ eV are undertaken,
following energy losses in the CMB and FIRB \cite{ss99}
and deflections in extragalactic magnetic fields (EGMF) for all particles
including secondary nuclei arising from photodisintegration.
We consider EGMF models that trace large-scale structure \cite{asm05},
as well as the case of negligible EGMF, although Galactic fields \cite{kst07} are not included.
A fraction $f_{\rm CR}$ of the accretion luminosity $L_{\rm acc}$ is converted to cosmic rays
with energy distributions $\propto E^{-\alpha} \exp (-E/E_{\max})$,
and we set $E_{\max}/Z=5 \times 10^{18}$ eV, a fair approximation
to estimates for each species obtained 
by comparing timescales as in Fig.\ref{fig:tacc}.
For the elemental composition at injection, the He/p ratio is taken to be 0.042.
All heavier elements are assumed
to have the same relative abundances at fixed energy/nucleon
as that of Galactic CR sources at GeV energies \cite{all07},
and scaled with respect to protons by the metallicity $\zeta$ of the accreting gas.
We take $\zeta=0.2$ as suggested by both observations and theory
for the gas flowing in from large-scale filaments \cite{nic05}.
An additional factor $A^{\beta}$ for the injected abundance of nuclei with mass number $A$
is introduced to take account of possible enhancement of heavier nuclei
due to nonlinear modification of shock structure by CRs \cite{bk99},
which may possibly be stronger here than for Galactic CRs 
due to the acceleration to much higher energies.

\section{Results and Discussion}

\begin{figure}[htb!]
\begin{center}
\includegraphics [width=0.47\textwidth]{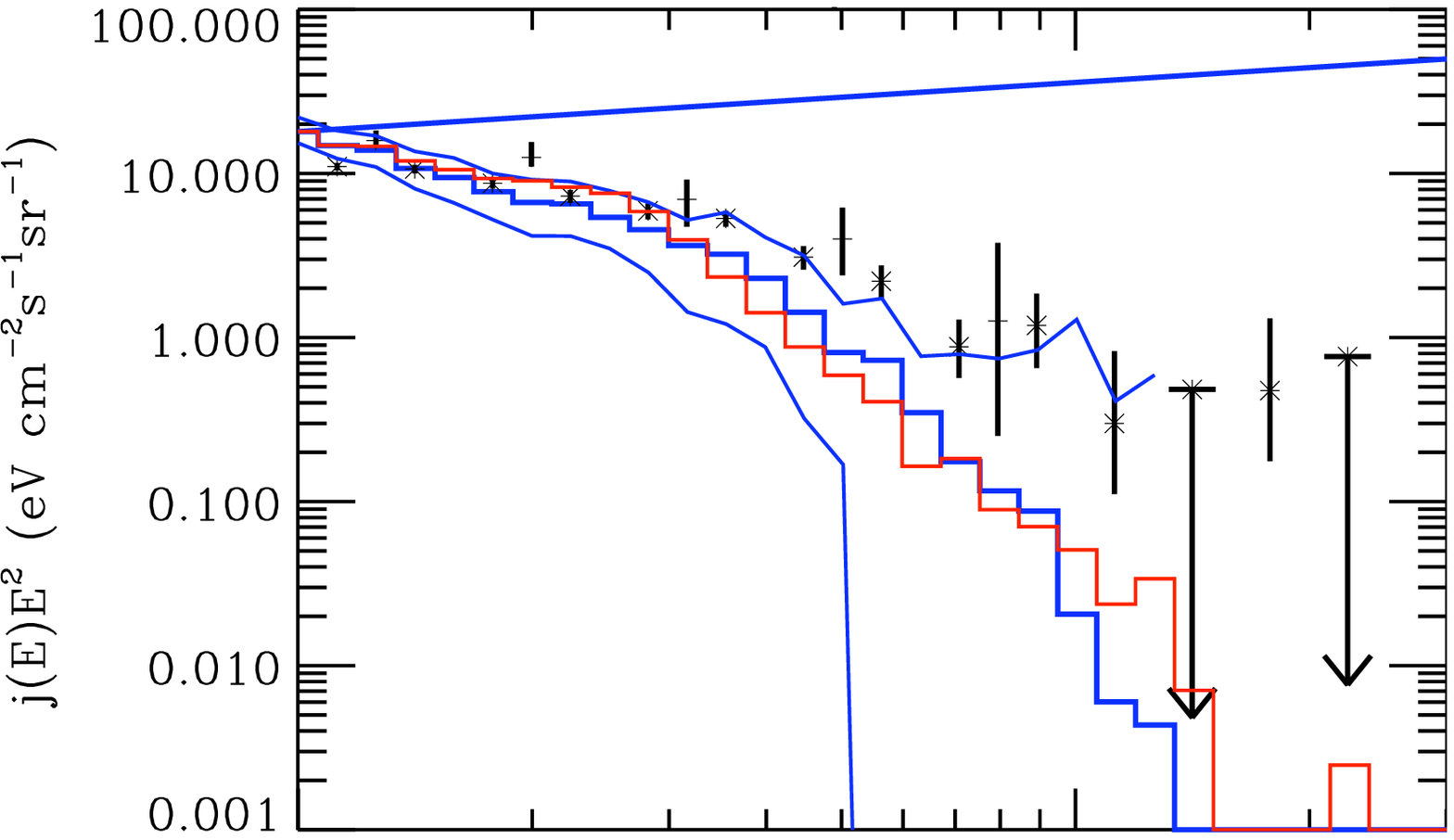}\\
\includegraphics [width=0.47\textwidth]{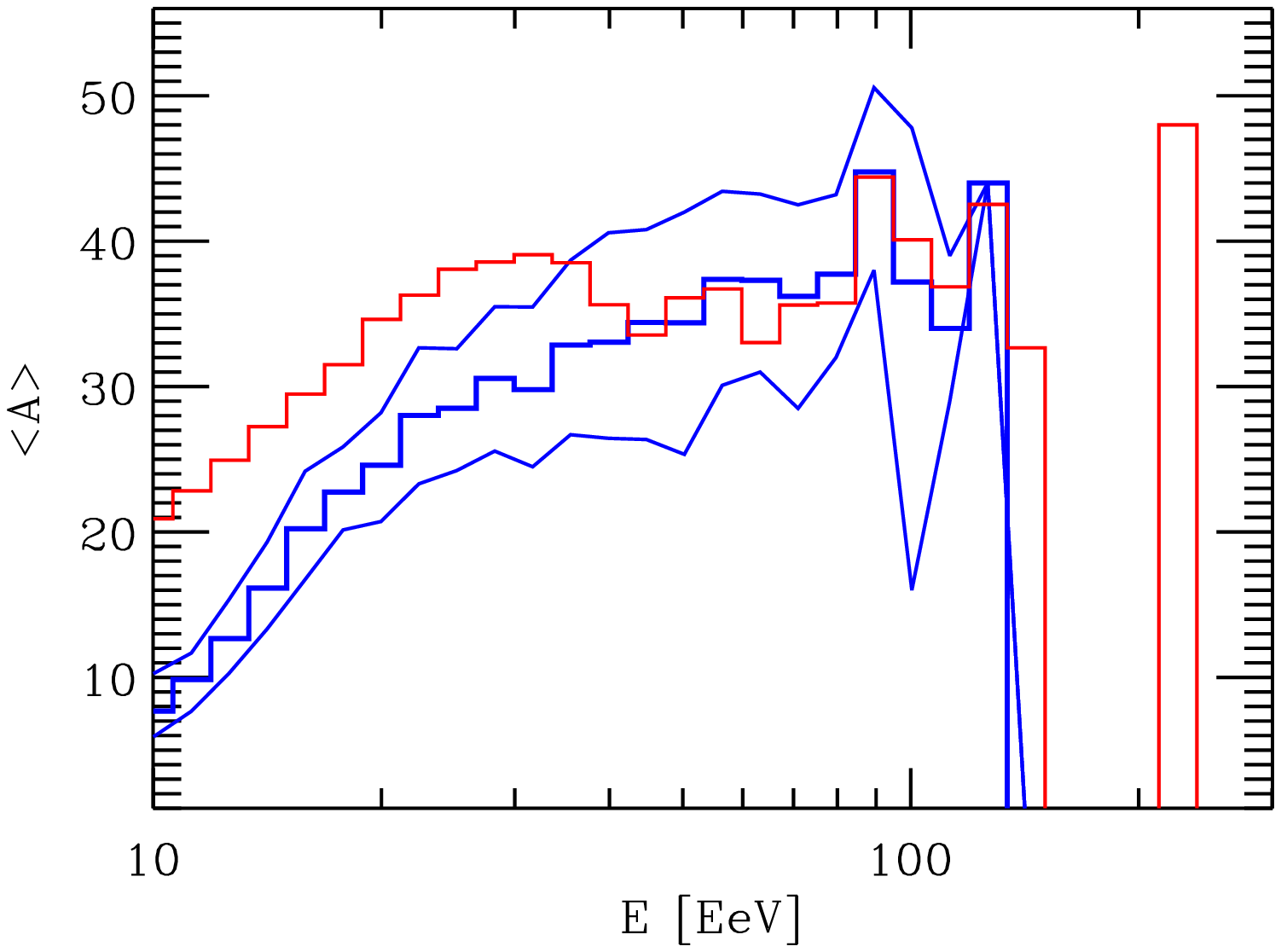}
\end{center}
\caption{Observed UHECR spectrum (top) and mean mass composition (bottom)
versus energy $E$ (1 EeV $\equiv 10^{18}$ eV)
from cluster accretion shocks for $\alpha=1.7$ and $\beta=0.5$, compared with the current data
for HiRes (bars) and Auger (asterisks).
The histograms are the average result over different model realizations
for the cases with (thick) and without (thin) EGMF,
and the thin curves outline the median deviations due to cosmic variance for the former case only.
The straight line in the top panel denotes the injection spectrum.}
\label{fig:speccomp}
\end{figure}

Fig.\ref{fig:speccomp} shows our results for the observed spectrum and composition
for $\alpha=1.7$ and $\beta=0.5$, which are consistent with the current data
for HiRes \cite{abb04} and Auger \cite{yam07} within about 2 sigma of combined
cosmic and statistical variance, for both the differential and integral fluxes
(and possibly AGASA \cite{tak98} as well).
Values of $\alpha<2$ are naturally expected at the high energy end
for nonlinear shock acceleration that accounts for the dynamical back reaction from CRs \cite{md01}.
The spectral steepening at $\ga 10^{20}$ eV is due both to propagation losses
and the $E_{\max}$ limit at the source.
Normalization to the observed flux and comparison with the available accretion power
for $M > 10^{15} M_\odot$ fixes $f_{\rm CR}$,
which is $\simeq 0.005-0.3$ for cases with EGMF and $\simeq 0.002$ for the case without.
Low values of $f_{CR}$ may reflect inefficient escape of CRs from the system,
which is conceivable in view of the converging nature of the accretion flow.
CR escape may be mediated by diffusion in directions away from the filaments,
or possibly by advection in outflows during merging events.

The mass composition at $\la 3 \times 10^{19}$ eV is predominantly light
and consistent with HiRes reports \cite{abb05},
while the rapid increase of the average mass at higher energies
is a clear prediction of the scenario to be tested by the new generation experiments
(and is in line with the latest Auger results on the shower elongation rate \cite{ung07}).

\begin{figure}[htb!]
\begin{center}
\includegraphics [width=0.42\textwidth]{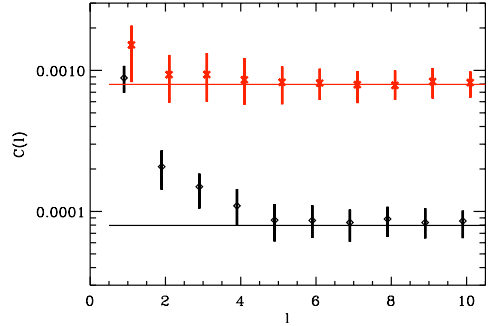}
\end{center}
\caption{Angular power spectrum $C(l)$ of UHECR arrival directions above $4 \times 10^{19}$ eV
versus multipole $l$,
for a realization with EGMF and a single, dominant cluster at $D \sim$ 50 Mpc.
The crosses are for 100 events with AGASA + SUGAR exposure
and diamonds for 1000 events with Auger North + South exposure.
Vertical bars indicate statistical errors.}
\label{fig:aniso}
\end{figure}

Despite the relative rarity of massive clusters in the local universe,
strong deflections of the highly charged nuclei in EGMF
allow consistency with the currently observed global isotropy (Fig. \ref{fig:aniso}).
On the other hand, with a sufficient number of accumulated events,
clear anistropies toward a small number of individual sources should appear,
although this prediction is subject to uncertainties in the EGMF and Galactic fields.

An aspect of this scenario that warrants further study is the spectral domain $< 10^{19}$ eV
and the implications for the Galactic-extragalactic transition region \cite{all07}.

\section{X-ray and Gamma-ray Signatures}

\begin{figure}
\begin{center}
\includegraphics [width=0.48\textwidth]{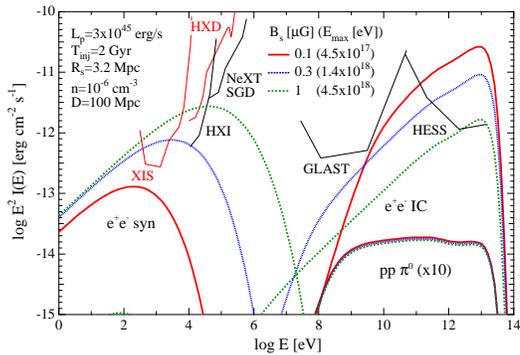}
\end{center}
\caption{Spectra of UHE proton-induced photopair emission from the accretion shock
of a Coma-like cluster, for $B_s =$0.1, 0.3 and 1 $\mu$G.
The sensitivities for a 1 degree extended source are overlayed for HESS, GLAST,
Suzaku XIS+HXD, and NeXT HXI+SGD.}
\label{fig:uhexg}
\end{figure}

If cluster accretion shocks are indeed accelerators of UHE particles,
we may look forward to very unique X-ray and gamma-ray emission
that can serve as valuable multimessenger signals.
Protons accelerated to $10^{18}$-$10^{19}$ eV in cluster accretion shocks
should efficiently channel energy into pairs of energy $10^{15}$-$10^{16}$ eV
through interactions with the CMB,
which then emit synchrotron radiation peaking in hard X-rays
and inverse Compton radiation in TeV gamma-rays.
Fig.\ref{fig:uhexg} displays the predicted spectra for a Coma-like cluster,
conservatively assuming that UHE proton injection continued only for a dynamical time $\simeq 2$ Gyr
(see Ref. \cite{ias05} for more details).
The detection prospects are promising
for Cerenkov telescopes such as HESS, VERITAS, CANGAROO III and MAGIC,
and hard X-ray observatories such as Suzaku and the future NeXT mission.
Photopair production by nuclei may also be efficient
and induce further interesting signals that are worth investigating.

Combined with such complementary information from X-ray and gamma-rays,
detailed measurements of UHECR composition and anisotropy
with facilities such as the Pierre Auger Observatory,
the Telescope Array, and the future Extreme Universe Space Observatory
should provide a clear test of whether the largest bound structures in the universe
are also the largest and most powerful particle accelerators.


\begin{thebibliography}{99}
\bibitem{nw00} M. Nagano and A. A. Watson, {\em Rev. Mod. Phys.} {\bf 72}, 689 (2000);
   J. W. Cronin, {\em Nucl. Phys. B Proc. Suppl.} {\bf 138}, 465 (2005).
\bibitem{hil84} A. M. Hillas, {\em Ann. Rev. Astron. Astrophys.} {\bf 22}, 425 (1984).
\bibitem{ino07} S. Inoue, arXiv:astro-ph/0701835.
\bibitem{ryu03} F. Miniati et al., {\em Astrophys. J.} {\bf 542}, 608 (2000);
D. Ryu et al.,  {\em Astrophys. J.} {\bf 593}, 599 (2003).
\bibitem{nma95} C. A. Norman, D. B. Melrose and A. Achterberg, {\em Astrophys. J.} {\bf 454}, 60 (1995).
\bibitem{isma07} S. Inoue, G. Sigl, F. Miniati and E. Armengaud, {\em Phys. Rev. D}, submitted (arXiv:astro-ph/0701167).
\bibitem{kwl04} U. Keshet, E. Waxman and A. Loeb, {\em Astrophys. J.} {\bf 617}, 281 (2004);
  V. Pavlidou and B. D. Fields,  {\em Astrophys. J.} {\bf 642}, 734 (2006).
\bibitem{rb02} T. H. Reiprich and H. B\"ohringer, {\em Astrophys. J.} {\bf 567}, 716 (2002).
\bibitem{krj96}
  H. Kang, D. Ryu and T. W. Jones, {\em Astrophys. J.} {\bf 456}, 422 (1996);
  H. Kang, J. P. Rachen, P. L. Biermann, {\em Mon. Not. R. A. S.} {\bf 286}, 257 (1997);
  M. Ostrowski and G. Siemieniec-Ozi\c{e}b\l o, {\em Astron. Astrophys.} {\bf 386}, 829 (2002).
\bibitem{ias05} S. Inoue, F. A. Aharonian and N. Sugiyama, {\em Astrophys. J.} {\bf 628}, L9 (2005).
\bibitem{fn06} L. Feretti and D. M. Neumann, {\em Astron. Astrophys.} {\bf 450}, L21 (2006);
  M. Johnston-Hollitt and R. Ekers, astro-ph/0411045.
\bibitem{bel04} A. R. Bell, {\em Mon. Not. R. A. S.} {\bf 353}, 550 (2004).
\bibitem{par06} E. Parizot et al., {\em Astron. Astrophys.} {\bf 453}, 387 (2006).
\bibitem{ss99}
  F. W. Stecker and M. H. Salamon, {\em Astrophys. J.} {\bf 512}, 521 (1999).
\bibitem{rb93} J. P. Rachen and P. L. Biermann, {\em Astron. Astrophys.} {\bf 272}, 161 (1993).
\bibitem{asm05} E. Armengaud, G. Sigl and F. Miniati, {\em Phys. Rev.} D {\bf 72}, 043009 (2005).
\bibitem{kst07}
  M. Kachelrie\ss, P. D. Serpico and M. Teshima, {\em Astropart. Phys.} {\bf 26}, 378 (2007).
\bibitem{all07}
  D. Allard, E. Parizot and A. V. Olinto, {\em Astropart. Phys.} {\bf 27}, 61 (2007)
\bibitem{nic05} F. Nicastro et al. {\em Nature} {\bf 433}, 495 (2005);
  R. Cen and J. P. Ostriker, {\em Astrophys. J.} {\bf 650}, 560 (2006)
\bibitem{bk99} E. G. Berezhko and L. T. Ksenofontov, {\em JETP} {\bf 89}, 391 (1999);
  L. O'C. Drury et al., {\em Sp. Sci. Rev.} {\bf 99}, 329 (2001).
\bibitem{abb04} R. U. Abbasi et al., {\em Phys. Rev. Lett.} {\bf 92}, 151101 (2004).
\bibitem{yam07} T. Yamamoto [Pierre Auger collaboration], 30th ICRC, arXiv:0707.2638.
\bibitem{tak98} M. Takeda et al., {\em Phys. Rev. Lett.} {\bf 81}, 1163 (1998).
\bibitem{md01} M. A. Malkov and L. O'C. Drury, {\em Rep. Prog. Phys.} {\bf 64}, 429 (2001).
\bibitem{abb05} R. U. Abbasi et al., {\em Astrophys. J.} {\bf 622}, 910 (2005).
\bibitem{ung07} M. Unger [Pierre Auger collaboration], these proceedings [arXiv:0706.1495]

\end{thebibliography}
\end{document}